# Inherent Symmetry and Microstructure Ambiguity in Micromechanics

*Kuang C. Liu and Anindya Ghoshal*
*U.S. Army Research Laboratory*
*Aberdeen Proving Ground, MD 21005*

## Abstract

The computational cost of micromechanics for heterogeneous materials can be reduced in certain cases where symmetric boundary conditions are applicable. We derived an eighth symmetric formulation of the Generalized Method of Cells for triply periodic microstructures. During this endeavor, an inherent symmetry was discovered. This implied that all repeating unit cells may be quarter symmetric representations of other microstructures. Additionally, it was discovered that a repeating unit cell can have columns of subcells swapped with no changes to the local or global fields. We concluded that first-order micromechanics are not well suited for capturing detailed or complex microstructures; however, higher-order theories, such as High Fidelity Generalized Method of Cells, can adequately model these microstructures.

## Introduction

The typical circular fiber in a square array, or sphere in a cubic array, is an ideal microstructure to apply symmetric boundary conditions. Several nongeneral micromechanics theories, which assume the circular fiber in square or hexagonal array, have already applied quarter symmetry to their repeating unit cells (RUCs) (Goldberg, 2001; Sun and Chen, 1991; for more detailed see Pindera et al. 2009) for simplification of the governing equations and the resulting computational efficiency. As the Generalized Method of Cells (GMC) (Paley and Aboudi, 1992; Aboudi, 1995) is suited to handle generalized doubly and triply period microstructures, there has not been a need to pursue the symmetry conditions for speed improvement. However, with the advent of the Multiscale Generalized Method of Cells (Liu et al., 2011; Liu, Arnold, and Chattopadhyay, 2011), which is a concurrent hierarchical implementation of the GMC to model multiple length scales, small improvements at one length scale can yield a large improvement overall. A reformulation of the GMC was performed by Pindera and Bednarcyk (1999), which significantly reduced the unknowns through relating local tractions instead of strains to global. The reformulation did not take advantage of geometric symmetry in the microstructure repeating unit cell, which can further improve speed. During an endeavor to apply symmetric boundary conditions in the GMC theory, the author observed inherent symmetry in the repeating unit cell. The inherent symmetry implies that the periodic repeating unit cell in the GMC is at all times representing only a fourth (doubly periodic) or eighth (triply periodic) of the actual microstructure. This implies two important considerations. First, no reformulation is necessary to model a quarter fiber, and secondly, the representation of complex microstructures maybe misleading. This paper first proves theoretically the inherent quarter symmetry and eighth symmetry in doubly and triply periodic assumptions, respectively. Next, we demonstrate the numerical proof to support the theory. Lastly, we discuss the implications of the inherent quarter symmetry. Although, the original formulation of GMC is used here, similar concepts can be applied to the reformulated versions also.

## Generalized Method of Cells with Symmetric Boundary Conditions

The following kinematics follow the approach of Aboudi (1995) using the same kinematic formulation until the application of symmetric conditions. The nomenclature is also preserved for comparison. We begin with a triply periodic microstructure of size $d \times h \times l$ whose periodic repeating until has symmetry about three planes, shown in Figure 1. The microstructure is discretized into $N_\alpha \times N_\beta \times N_\gamma$ rectangular cuboids, where $N_\alpha$, $N_\beta$, and $N_\gamma$ are even numbers. Each cuboid or subcell is aligned with the global $x_1$, $x_2$, and $x_3$ coordinate system shown in Figure 1 and can be referenced by a set of indices, $(\alpha\beta\gamma)$, each corresponding to a unique location their respective axes. The planes of symmetry lie at the interfaces between $\alpha = N_\alpha/2$ and $\alpha = N_\alpha/2 + 1$, $\beta = N_\beta/2$ and $\beta = N_\beta/2 + 1$, and $\gamma = N_\gamma/2$ and $\gamma = N_\gamma/2 + 1$. Each subcell $(\alpha\beta\gamma)$ has a volume given by $d_\alpha \times h_\beta \times l_\gamma$ and local coordinates $\bar{x}_1^{(\alpha)}$, $\bar{x}_2^{(\beta)}$, and $\bar{x}_3^{(\gamma)}$ that are aligned with the correspond global coordinates. A first-order expansion of the displacement field in a subcell in terms the distances from the center of each subcell, i.e., $\bar{x}_1^{(\alpha)}$, $\bar{x}_2^{(\beta)}$, and $\bar{x}_3^{(\gamma)}$, can be written as

$$u_i^{(\alpha\beta\gamma)} = w_i^{(\alpha\beta\gamma)}(\mathbf{x}) + \bar{x}_1^{(\alpha)} \phi_i^{(\alpha\beta\gamma)} + \bar{x}_2^{(\beta)} \chi_i^{(\alpha\beta\gamma)} + \bar{x}_3^{(\gamma)} \psi_i^{(\alpha\beta\gamma)} \quad i = 1, 2, 3. \tag{1}$$

Here, $w_i^{(\alpha\beta\gamma)}$ are the displacements at the center of the subcell and the variables $\phi_i^{(\alpha\beta\gamma)}$, $\chi_i^{(\alpha\beta\gamma)}$, and $\psi_i^{(\alpha\beta\gamma)}$ are microvariables for the first-order expansion about the local coordinates $\bar{x}_1^{(\alpha)}$, $\bar{x}_2^{(\beta)}$, and $\bar{x}_3^{(\gamma)}$. The variable, $\mathbf{x} = (x_1, x_2, x_3)$, is the center location of a subcell with respect to the fixed global coordinate system. By applying infinitesimal strain theory, the small strain tensor in a subcell can be related to the displacement field by

$$\varepsilon_{ij}^{(\alpha\beta\gamma)} = \frac{1}{2}\left(u_{i,j}^{(\alpha\beta\gamma)} + u_{j,i}^{(\alpha\beta\gamma)}\right) \quad i,j = 1, 2, 3 \tag{2}$$

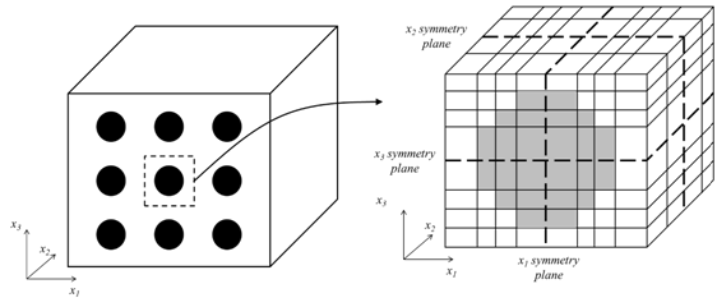

Figure 1. A repeating unit cell of a circular fiber in a periodic array is shown discretized into orthogonal cuboid subcells. Although the discretization in the $x_2$ direction is not necessary, it is used to maintain a general symmetric formulation.

where $,_1^{(\alpha\beta\gamma)} = \dfrac{\partial}{\partial \overline{x}_1^{(\alpha)}}$, $,_2^{(\alpha\beta\gamma)} = \dfrac{\partial}{\partial \overline{x}_2^{(\beta)}}$, and $,_3^{(\alpha\beta\gamma)} = \dfrac{\partial}{\partial \overline{x}_3^{(\gamma)}}$.

Therefore, each strain component can then be computed in terms of the microvariables. Due to the first-order expansion of the displacement field, this results in constant strains within the subcell, which are referred to as average (denoted by the overbar) strains.

$$\overline{\varepsilon}_{11}^{(\alpha\beta\gamma)} = \phi_1^{(\alpha\beta\gamma)}$$
$$\overline{\varepsilon}_{22}^{(\alpha\beta\gamma)} = \chi_2^{(\alpha\beta\gamma)}$$
$$\overline{\varepsilon}_{33}^{(\alpha\beta\gamma)} = \psi_3^{(\alpha\beta\gamma)}$$
$$\overline{\gamma}_{23}^{(\alpha\beta\gamma)} = 2\overline{\varepsilon}_{23}^{(\alpha\beta\gamma)} = \chi_3^{(\alpha\beta\gamma)} + \psi_2^{(\alpha\beta\gamma)}$$
$$\overline{\gamma}_{13}^{(\alpha\beta\gamma)} = 2\overline{\varepsilon}_{13}^{(\alpha\beta\gamma)} = \phi_3^{(\alpha\beta\gamma)} + \psi_1^{(\alpha\beta\gamma)}$$
$$\overline{\gamma}_{12}^{(\alpha\beta\gamma)} = 2\overline{\varepsilon}_{12}^{(\alpha\beta\gamma)} = \phi_2^{(\alpha\beta\gamma)} + \chi_1^{(\alpha\beta\gamma)}$$

(3)

The average strains in the composite RUC can be written as

$$\overline{\varepsilon}_{ij} = \frac{1}{d}\frac{1}{h}\frac{1}{l}\sum_{\alpha=1}^{N_\alpha}\sum_{\beta=1}^{N_\beta}\sum_{\gamma=1}^{N_\gamma}\overline{\varepsilon}_{ij}^{(\alpha\beta\gamma)} d_\alpha h_\beta l_\gamma .$$

(4)

Assuming a thermo-elastoplastic constitutive model, the stress-strain constitutive relationship can be used to determine the average subcell stresses, i.e.,

$$\overline{\sigma}_{ij}^{(\alpha\beta\gamma)} = C_{ijkl}^{(\alpha\beta\gamma)}\left(\overline{\varepsilon}_{kl}^{(\alpha\beta\gamma)} - \overline{\varepsilon}_{kl}^{I(\alpha\beta\gamma)} - \overline{\varepsilon}_{kl}^{T(\alpha\beta\gamma)}\right),$$

(5)

where $\overline{\sigma}_{ij}^{(\alpha\beta\gamma)}$ is the average stress tensor, $C_{ijkl}^{(\alpha\beta\gamma)}$ is the elastic stiffness tensor, $\overline{\varepsilon}_{kl}^{I(\alpha\beta\gamma)}$ is the inelastic strains, and $\overline{\varepsilon}_{kl}^{T(\alpha\beta\gamma)}$ is the thermal strains. The global average stress can be defined in the same manner as the strains by

$$\overline{\sigma}_{ij} = \frac{1}{d}\frac{1}{h}\frac{1}{l}\sum_{\alpha=1}^{N_\alpha}\sum_{\beta=1}^{N_\beta}\sum_{\gamma=1}^{N_\gamma}\overline{\sigma}_{ij}^{(\alpha\beta\gamma)} d_\alpha h_\beta l_\gamma .$$

(6)

In order to solve for the microvariables, a set of interfacial boundary conditions for continuity of traction and displacement must be established. For each subcell, the neighboring subcell must have an equivalent set of displacement components at the interface. This leads to the following set of conditions,

$$u_i^{(\alpha\beta\gamma)}\bigg|_{\overline{x}_1^\alpha = d_\alpha/2} = u_i^{(\hat{\alpha}\beta\gamma)}\bigg|_{\overline{x}_1^{\hat{\alpha}} = -d_{\hat{\alpha}}/2}$$
$$u_i^{(\alpha\beta\gamma)}\bigg|_{\overline{x}_2^\beta = h_\beta/2} = u_i^{(\alpha\hat{\beta}\gamma)}\bigg|_{\overline{x}_2^{\hat{\beta}} = -h_{\hat{\beta}}/2}.$$
$$u_i^{(\alpha\beta\gamma)}\bigg|_{\overline{x}_3^\gamma = l_\gamma/2} = u_i^{(\alpha\beta\hat{\gamma})}\bigg|_{\overline{x}_3^{\hat{\gamma}} = -l_{\hat{\gamma}}/2}$$

(7)

applied for $\alpha = 1,...,N_\alpha$, $\beta = 1,...,N_\beta$, and $\gamma = 1,...,N_\gamma$, where the ^ denotes an adjacent subcell. In the GMC, these continuity conditions are applied in an average sense across the boundary yielding the following conditions

$$\int_{-l_\gamma/2}^{l_\gamma/2}\int_{-h_\beta/2}^{h_\beta/2} u_i^{(\alpha\beta\gamma)}\bigg|_{\bar{x}_1^\alpha = d_\alpha/2} d\bar{x}_2^\beta d\bar{x}_3^\gamma = \int_{-l_\gamma/2}^{l_\gamma/2}\int_{-h_\beta/2}^{h_\beta/2} u_i^{(\hat{\alpha}\beta\gamma)}\bigg|_{\bar{x}_1^{\hat{\alpha}} = -d_{\hat{\alpha}}/2} d\bar{x}_2^\beta d\bar{x}_3^\gamma \tag{8}$$

$$\int_{-l_\gamma/2}^{l_\gamma/2}\int_{-d_\alpha/2}^{d_\alpha/2} u_i^{(\alpha\beta\gamma)}\bigg|_{\bar{x}_2^\beta = h_\beta/2} d\bar{x}_1^\alpha d\bar{x}_3^\gamma = \int_{-l_\gamma/2}^{l_\gamma/2}\int_{-d_\alpha/2}^{d_\alpha/2} u_i^{(\alpha\hat{\beta}\gamma)}\bigg|_{\bar{x}_2^{\hat{\beta}} = -h_{\hat{\beta}}/2} d\bar{x}_1^\alpha d\bar{x}_3^\gamma$$

$$\int_{-h_\beta/2}^{h_\beta/2}\int_{-d_\alpha/2}^{d_\alpha/2} u_i^{(\alpha\beta\gamma)}\bigg|_{\bar{x}_3^\gamma = l_\gamma/2} d\bar{x}_1^\alpha d\bar{x}_2^\beta = \int_{-h_\beta/2}^{h_\beta/2}\int_{-d_\alpha/2}^{d_\alpha/2} u_i^{(\alpha\beta\hat{\gamma})}\bigg|_{\bar{x}_3^{\hat{\gamma}} = -l_{\hat{\gamma}}/2} d\bar{x}_1^\alpha d\bar{x}_2^\beta .$$

Substitution of the displacement field expansion into the above equation yields a set of equations in terms of the microvariables,

$$w_i^{(\alpha\beta\gamma)} + \frac{d_\alpha}{2}\phi_i^{(\alpha\beta\gamma)} = w_i^{(\hat{\alpha}\beta\gamma)} - \frac{d_{\hat{\alpha}}}{2}\phi_i^{(\hat{\alpha}\beta\gamma)} \tag{9}$$

$$w_i^{(\alpha\beta\gamma)} + \frac{h_\beta}{2}\chi_i^{(\alpha\beta\gamma)} = w_i^{(\alpha\hat{\beta}\gamma)} - \frac{h_{\hat{\beta}}}{2}\chi_i^{(\alpha\hat{\beta}\gamma)}$$

$$w_i^{(\alpha\beta\gamma)} + \frac{l_\gamma}{2}\psi_i^{(\alpha\beta\gamma)} = w_i^{(\alpha\beta\hat{\gamma})} - \frac{l_{\hat{\gamma}}}{2}\psi_i^{(\alpha\beta\hat{\gamma})}.$$

In the above equation, all the field variables, $w_i$, are evaluated at the center of the subcell; however, it is necessary to evaluate these at a common location, the interface. In the global coordinate system, the interface is defined as

$$x_1^I = \left(x_1^{(\alpha)} + \frac{d_\alpha}{2}, x_2, x_3\right) = \left(x_1^{(\hat{\alpha})} - \frac{d_{\hat{\alpha}}}{2}, x_2, x_3\right), \tag{10}$$

$$x_2^I = \left(x_1, x_2^{(\beta)} + \frac{h_\beta}{2}, x_3\right) = \left(x_1, x_2^{(\hat{\beta})} - \frac{h_{\hat{\beta}}}{2}, x_3\right),$$

and

$$x_3^I = \left(x_1, x_2, x_3^{(\gamma)} + \frac{l_\gamma}{2}\right) = \left(x_1, x_2, x_3^{(\hat{\gamma})} - \frac{l_{\hat{\gamma}}}{2}\right).$$

To evaluate the field variables ($w_i^{(\alpha\beta\gamma)}$) at the interface, a first-order Taylor expansion about the common interface is used. The continuity conditions then become

$$w_i^{(\alpha\beta\gamma)} - \frac{d_\alpha}{2}\left(\frac{\partial w_i^{(\alpha\beta\gamma)}}{\partial x_1} - \phi_i^{(\alpha\beta\gamma)}\right) = w_i^{(\hat{\alpha}\beta\gamma)} + \frac{d_{\hat{\alpha}}}{2}\left(\frac{\partial w_i^{(\hat{\alpha}\beta\gamma)}}{\partial x_1} - \phi_i^{(\hat{\alpha}\beta\gamma)}\right) \tag{11}$$

$$w_i^{(\alpha\beta\gamma)} - \frac{h_\beta}{2}\left(\frac{\partial w_i^{(\alpha\beta\gamma)}}{\partial x_2} - \chi_i^{(\alpha\beta\gamma)}\right) = w_i^{(\alpha\hat{\beta}\gamma)} + \frac{h_{\hat{\beta}}}{2}\left(\frac{\partial w_i^{(\alpha\hat{\beta}\gamma)}}{\partial x_2} - \chi_i^{(\alpha\hat{\beta}\gamma)}\right)$$

$$w_i^{(\alpha\beta\gamma)} - \frac{l_\gamma}{2}\left(\frac{\partial w_i^{(\alpha\beta\gamma)}}{\partial x_3} - \psi_i^{(\alpha\beta\gamma)}\right) = w_i^{(\alpha\beta\hat{\gamma})} + \frac{l_{\hat{\gamma}}}{2}\left(\frac{\partial w_i^{(\alpha\beta\hat{\gamma})}}{\partial x_3} - \psi_i^{(\alpha\beta\hat{\gamma})}\right),$$

where each field variable and field variable derivative is evaluated at the interface. Next, let the functions $F$, $G$, and $H$ be defined as

$$F_i^{(\alpha)} = w_i^{(\alpha\beta\gamma)}\Big|_{\mathbf{x}=x_1^I} + f_i^{(\alpha)} - w_i^{(\hat{\alpha}\beta\gamma)}\Big|_{\mathbf{x}=x_1^I} - f_i^{(\hat{\alpha})} \tag{12}$$

$$G_i^{(\beta)} = w_i^{(\alpha\beta\gamma)}\Big|_{\mathbf{x}=x_2^I} + g_i^{(\beta)} - w_i^{(\alpha\hat{\beta}\gamma)}\Big|_{\mathbf{x}=x_2^I} - g_i^{(\hat{\beta})}$$

$$H_i^{(\gamma)} = w_i^{(\alpha\beta\gamma)}\Big|_{\mathbf{x}=x_3^I} + h_i^{(\gamma)} - w_i^{(\alpha\beta\hat{\gamma})}\Big|_{\mathbf{x}=x_3^I} - h_i^{(\hat{\gamma})}$$

where

$$f_i^{(\alpha)} = -\frac{d_\alpha}{2}\left(\frac{\partial w_i^{(\alpha\beta\gamma)}}{\partial x_1}\Big|_{\mathbf{x}=x_1^I} - \phi_i^{(\alpha\beta\gamma)}\right) \tag{13}$$

$$g_i^{(\beta)} = -\frac{h_\beta}{2}\left(\frac{\partial w_i^{(\alpha\beta\gamma)}}{\partial x_2}\Big|_{\mathbf{x}=x_2^I} - \chi_i^{(\alpha\beta\gamma)}\right)$$

$$h_i^{(\gamma)} = -\frac{l_\gamma}{2}\left(\frac{\partial w_i^{(\alpha\beta\gamma)}}{\partial x_3}\Big|_{\mathbf{x}=x_3^I} - \psi_i^{(\alpha\beta\gamma)}\right).$$

The three continuity equations can then be rewritten as

$$F_i^{(\alpha)} = 0 \quad \alpha = 1,...,N_\alpha \tag{14}$$
$$G_i^{(\beta)} = 0 \quad \beta = 1,...,N_\beta$$
$$H_i^{(\gamma)} = 0 \quad \gamma = 1,...,N_\gamma$$

and subsequently these can be written as a summation series

$$\sum_{\alpha=1}^{N_\alpha} F_i^{(\alpha)} = 0, \quad \sum_{\beta=1}^{N_\beta} G_i^{(\beta)} = 0, \quad \sum_{\gamma=1}^{N_\gamma} H_i^{(\gamma)} = 0. \tag{15}$$

These summations lead to the conclusion that

$$\sum_{\alpha=1}^{N_\alpha} f_i^{(\alpha)} = 0, \quad \sum_{\beta=1}^{N_\beta} g_i^{(\beta)} = 0, \quad \sum_{\gamma=1}^{N_\gamma} h_i^{(\gamma)} = 0. \tag{16}$$

Under first-order theory, in which the second derivative of $w_i^{(\alpha\beta\gamma)}$ is zero,

$$\frac{\partial f_i^{(\alpha)}}{\partial x_1} = 0, \quad \frac{\partial g_i^{(\beta)}}{\partial x_2} = 0, \quad \frac{\partial h_i^{(\gamma)}}{\partial x_3} = 0. \tag{17}$$

In addition, differentiation of the continuity equations with respect to $x_1$, $x_2$, and $x_3$ results in

$$\frac{\partial w_i^{(\alpha\beta\gamma)}}{\partial x_1} = \frac{\partial w_i^{(\hat{\alpha}\beta\gamma)}}{\partial x_1} \tag{18}$$

$$\frac{\partial w_i^{(\alpha\beta\gamma)}}{\partial x_2} = \frac{\partial w_i^{(\alpha\hat{\beta}\gamma)}}{\partial x_2}$$

$$\frac{\partial w_i^{(\alpha\beta\gamma)}}{\partial x_3} = \frac{\partial w_i^{(\alpha\beta\hat{\gamma})}}{\partial x_3},$$

which can be satisfied by assuming that common displacement functions, $w_i$, exist such that

$$w_i^{(\alpha\beta\gamma)} = w_i \tag{19}$$

and therefore

$$w_i^{(\alpha\beta\gamma)}\Big|_{\mathbf{x}=x_j^I} = w_i. \tag{20}$$

Using this assumption and Eq. 16 a set of continuum relations can be derived

$$\sum_{\alpha=1}^{N_\alpha} d_\alpha \phi_i^{(\alpha\beta\gamma)} = d \frac{\partial w_i}{\partial x_1} \tag{21a-c}$$

$$\sum_{\beta=1}^{N_\beta} h_\beta \chi_i^{(\alpha\beta\gamma)} = h \frac{\partial w_i}{\partial x_2}$$

$$\sum_{\gamma=1}^{N_\gamma} l_\gamma \psi_i^{(\alpha\beta\gamma)} = l \frac{\partial w_i}{\partial x_3}.$$

The previously defined small strain tensor can be written in terms of the common displacement functions,

$$\bar{\varepsilon}_{ij} = \frac{1}{2}\left(w_{i,j} + w_{j,i}\right). \tag{22}$$

Substitution of this into the set of continuum relations yields

$$\sum_{\alpha=1}^{N_\alpha} d_\alpha \bar{\varepsilon}_{11}^{(\alpha\beta\gamma)} = d\bar{\varepsilon}_{11}, \quad \beta = 1,...,N_\beta, \gamma = 1,...,N_\gamma \tag{23a-c}$$

$$\sum_{\beta=1}^{N_\beta} h_\beta \bar{\varepsilon}_{22}^{(\alpha\beta\gamma)} = h\bar{\varepsilon}_{22}, \quad \alpha = 1,...,N_\alpha, \gamma = 1,...,N_\gamma$$

$$\sum_{\gamma=1}^{N_\gamma} l_\gamma \bar{\varepsilon}_{33}^{(\alpha\beta\gamma)} = l\bar{\varepsilon}_{33}, \quad \alpha = 1,...,N_\alpha, \beta = 1,...,N_\beta.$$

Combining Eq. 21a multiplied by $h_\beta$ summed over $\beta$ for $i=1$ with Eq. 21b multiplied by $d_\alpha$ summed over $\alpha$ for $i=2$ yields

$$\sum_{\alpha=1}^{N_\alpha}\sum_{\beta=1}^{N_\beta} d_\alpha h_\beta \left(\phi_2^{(\alpha\beta\gamma)} + \chi_1^{(\alpha\beta\gamma)}\right) = dh\left(\frac{\partial w_1}{\partial x_2} + \frac{\partial w_2}{\partial x_1}\right), \quad \gamma = 1,...,N_\gamma \tag{24a,b}$$

$$\sum_{\alpha=1}^{N_\alpha}\sum_{\beta=1}^{N_\beta} d_\alpha h_\beta \bar{\varepsilon}_{12}^{(\alpha\beta\gamma)} = dh\bar{\varepsilon}_{12}, \quad \gamma = 1,...,N_\gamma.$$

Similar operations yield

$$\sum_{\beta=1}^{N_\beta}\sum_{\gamma=1}^{N_\gamma} h_\beta l_\gamma \bar{\varepsilon}_{23}^{(\alpha\beta\gamma)} = hl\bar{\varepsilon}_{23}, \quad \alpha = 1,...,N_\alpha \tag{25a,b}$$

$$\sum_{\alpha=1}^{N_\alpha}\sum_{\gamma=1}^{N_\gamma} d_\alpha l_\gamma \bar{\varepsilon}_{13}^{(\alpha\beta\gamma)} = dl\bar{\varepsilon}_{13}, \quad \beta = 1,...,N_\beta.$$

Now, we can apply the strain conditions that arise from symmetry. The symmetry is applied by equating strains at subcells on opposite sides of the symmetry plane, i.e.,

$$\bar{\varepsilon}_{ij}^{\alpha\beta\gamma} = \bar{\varepsilon}_{ij}^{\tilde{\alpha}\beta\gamma} = \bar{\varepsilon}_{ij}^{\alpha\tilde{\beta}\gamma} = \bar{\varepsilon}_{ij}^{\alpha\beta\tilde{\gamma}} = \bar{\varepsilon}_{ij}^{\tilde{\alpha}\tilde{\beta}\gamma} = \bar{\varepsilon}_{ij}^{\tilde{\alpha}\beta\tilde{\gamma}} = \bar{\varepsilon}_{ij}^{\alpha\tilde{\beta}\tilde{\gamma}} = \bar{\varepsilon}_{ij}^{\tilde{\alpha}\tilde{\beta}\tilde{\gamma}}, \tag{26}$$

where

$$\alpha = 1 : N_\alpha / 2, \tag{27}$$

$$\beta = 1 : N_\beta / 2,$$

and

$$\gamma = 1 : N_\gamma / 2,$$

and

$$\tilde{\alpha} = N_\alpha + 1 - \alpha, \tag{28}$$

$$\tilde{\beta} = N_\beta + 1 - \beta,$$

and

$$\tilde{\gamma} = N_\gamma + 1 - \gamma.$$

Substitution of the symmetry conditions into the continuity equations yields a new set of constraints,

$$2\sum_{\alpha=1}^{N_\alpha/2} d_\alpha \bar{\varepsilon}_{11}^{(\alpha\beta\gamma)} = d\bar{\varepsilon}_{11} \qquad \beta=1,...,N_\beta/2, \gamma=1,...,N_\gamma/2 \tag{29}$$

$$2\sum_{\beta=1}^{N_\beta/2} h_\beta \bar{\varepsilon}_{22}^{(\alpha\beta\gamma)} = h\bar{\varepsilon}_{22} \qquad \alpha=1,...,N_\alpha/2, \gamma=1,...,N_\gamma/2$$

$$2\sum_{\gamma=1}^{N_\gamma/2} l_\gamma \bar{\varepsilon}_{33}^{(\alpha\beta\gamma)} = l\bar{\varepsilon}_{33} \qquad \alpha=1,...,N_\alpha/2, \beta=1,...,N_\beta/2$$

$$4\sum_{\alpha=1}^{N_\alpha/2}\sum_{\beta=1}^{N_\beta/2} d_\alpha h_\beta \bar{\varepsilon}_{12}^{(\alpha\beta\gamma)} = dh\bar{\varepsilon}_{12} \qquad \gamma=1,...,N_\gamma/2$$

$$4\sum_{\beta=1}^{N_\beta/2}\sum_{\gamma=1}^{N_\gamma/2} h_\beta l_\gamma \bar{\varepsilon}_{23}^{(\alpha\beta\gamma)} = hl\bar{\varepsilon}_{23} \qquad \alpha=1,...,N_\alpha/2$$

$$4\sum_{\alpha=1}^{N_\alpha/2}\sum_{\gamma=1}^{N_\gamma/2} d_\alpha l_\gamma \bar{\varepsilon}_{13}^{(\alpha\beta\gamma)} = dl\bar{\varepsilon}_{13} \qquad \beta=1,...,N_\beta/2.$$

These global-local strain relationships can be cast into matrix form as

$$A_G \varepsilon_s = J\bar{\varepsilon}, \tag{30}$$

where

$$\bar{\varepsilon} = \left(\bar{\varepsilon}_{11}, \bar{\varepsilon}_{22}, \bar{\varepsilon}_{33}, 2\bar{\varepsilon}_{23}, 2\bar{\varepsilon}_{13}, 2\bar{\varepsilon}_{12}\right) \tag{31}$$

and

$$\varepsilon_s = \left(\bar{\varepsilon}^{(111)},...,\bar{\varepsilon}^{(N_\alpha/2N_\beta/2N_\gamma/2)}\right). \tag{32}$$

The interfacial traction continuity conditions, like the displacement continuity conditions, are also imposed on an average sense. The conditions can be expressed as

$$\bar{\sigma}_{1i}^{(\alpha\beta\gamma)} = \bar{\sigma}_{1i}^{(\hat{\alpha}\beta\gamma)} \tag{33}$$

$$\bar{\sigma}_{2i}^{(\alpha\beta\gamma)} = \bar{\sigma}_{2i}^{(\alpha\hat{\beta}\gamma)}$$

$$\bar{\sigma}_{3i}^{(\alpha\beta\gamma)} = \bar{\sigma}_{3i}^{(\alpha\beta\hat{\gamma})}$$

for $i,j,k=1,2,3$ and $\alpha=1,...N_\alpha$, $\beta=1,...,N_\beta$, and $\gamma=1,...,N_\gamma$. However, only a subset of these equations are independent and they can expressed as

$$\bar{\sigma}_{11}^{(\alpha\beta\gamma)} = \bar{\sigma}_{11}^{(\hat{\alpha}\beta\gamma)} \quad \alpha = 1,...,N_\alpha - 1, \beta = 1,...,N_\beta, \gamma = 1,...,N_\gamma \tag{34}$$

$$\bar{\sigma}_{22}^{(\alpha\beta\gamma)} = \bar{\sigma}_{22}^{(\alpha\hat{\beta}\gamma)} \quad \alpha = 1,...,N_\alpha, \beta = 1,...,N_\beta - 1, \gamma = 1,...,N_\gamma$$

$$\bar{\sigma}_{33}^{(\alpha\beta\gamma)} = \bar{\sigma}_{33}^{(\alpha\beta\hat{\gamma})} \quad \alpha = 1,...,N_\alpha, \beta = 1,...,N_\beta, \gamma = 1,...,N_\gamma - 1$$

$$\bar{\sigma}_{23}^{(\alpha\beta\gamma)} = \bar{\sigma}_{23}^{(\alpha\hat{\beta}\gamma)} \quad \alpha = 1,...,N_\alpha, \beta = 1,...,N_\beta - 1, \gamma = 1,...,N_\gamma$$

$$\bar{\sigma}_{32}^{(\alpha\beta\gamma)} = \bar{\sigma}_{32}^{(\alpha\beta\hat{\gamma})} \quad \alpha = 1,...,N_\alpha, \beta = N_\beta, \gamma = 1,...,N_\gamma - 1$$

$$\bar{\sigma}_{13}^{(\alpha\beta\gamma)} = \bar{\sigma}_{13}^{(\hat{\alpha}\beta\gamma)} \quad \alpha = 1,...,N_\alpha - 1, \beta = 1,...,N_\beta, \gamma = 1,...,N_\gamma$$

$$\bar{\sigma}_{31}^{(\alpha\beta\gamma)} = \bar{\sigma}_{31}^{(\alpha\beta\hat{\gamma})} \quad \alpha = N_\alpha, \beta = 1,...,N_\beta, \gamma = 1,...,N_\gamma - 1$$

$$\bar{\sigma}_{12}^{(\alpha\beta\gamma)} = \bar{\sigma}_{12}^{(\hat{\alpha}\beta\gamma)} \quad \alpha = 1,...,N_\alpha - 1, \beta = 1,...,N_\beta, \gamma = N_\gamma$$

$$\bar{\sigma}_{21}^{(\alpha\beta\gamma)} = \bar{\sigma}_{21}^{(\alpha\hat{\beta}\gamma)} \quad \alpha = N_\alpha - 1, \beta = 1,...,N_\beta - 1, \gamma = N_\gamma.$$

Here we apply the stress constraints that result from the symmetric boundary conditions,

$$\bar{\sigma}_{ij}^{\alpha\beta\gamma} = \bar{\sigma}_{ij}^{\tilde{\alpha}\beta\gamma} = \bar{\sigma}_{ij}^{\alpha\tilde{\beta}\gamma} = \bar{\sigma}_{ij}^{\alpha\beta\tilde{\gamma}} = \bar{\sigma}_{ij}^{\tilde{\alpha}\tilde{\beta}\gamma} = \bar{\sigma}_{ij}^{\tilde{\alpha}\beta\tilde{\gamma}} = \bar{\sigma}_{ij}^{\alpha\tilde{\beta}\tilde{\gamma}} = \bar{\sigma}_{ij}^{\tilde{\alpha}\tilde{\beta}\tilde{\gamma}}, \tag{35}$$

where the

$$\alpha = 1 : N_\alpha / 2, \tag{36}$$

$$\beta = 1 : N_\beta / 2,$$

and

$$\gamma = 1 : N_\gamma / 2,$$

and

$$\tilde{\alpha} = N_\alpha + 1 - \alpha, \tag{37}$$

$$\tilde{\beta} = N_\beta + 1 - \beta,$$

and

$$\tilde{\gamma} = N_\gamma + 1 - \gamma.$$

Substitution of these terms into the traction continuity equations results in

$$\bar{\sigma}_{11}^{(\alpha\beta\gamma)} = \bar{\sigma}_{11}^{(\hat{\alpha}\beta\gamma)} \quad \alpha = 1,...,N_\alpha/2-1, \beta = 1,...,N_\beta/2, \gamma = 1,...,N_\gamma/2 \quad (38)$$

$$\bar{\sigma}_{22}^{(\alpha\beta\gamma)} = \bar{\sigma}_{22}^{(\alpha\hat{\beta}\gamma)} \quad \alpha = 1,...,N_\alpha/2, \beta = 1,...,N_\beta/2-1, \gamma = 1,...,N_\gamma/2$$

$$\bar{\sigma}_{33}^{(\alpha\beta\gamma)} = \bar{\sigma}_{33}^{(\alpha\beta\hat{\gamma})} \quad \alpha = 1,...,N_\alpha/2, \beta = 1,...,N_\beta/2, \gamma = 1,...,N_\gamma/2-1$$

$$\bar{\sigma}_{23}^{(\alpha\beta\gamma)} = \bar{\sigma}_{23}^{(\alpha\hat{\beta}\gamma)} \quad \alpha = 1,...,N_\alpha/2, \beta = 1,...,N_\beta/2-1, \gamma = 1,...,N_\gamma/2$$

$$\bar{\sigma}_{32}^{(\alpha\beta\gamma)} = \bar{\sigma}_{32}^{(\alpha\beta\hat{\gamma})} \quad \alpha = 1,...,N_\alpha/2, \beta = N_\beta/2, \gamma = 1,...,N_\gamma/2-1$$

$$\bar{\sigma}_{13}^{(\alpha\beta\gamma)} = \bar{\sigma}_{13}^{(\hat{\alpha}\beta\gamma)} \quad \alpha = 1,...,N_\alpha/2-1, \beta = 1,...,N_\beta/2, \gamma = 1,...,N_\gamma/2$$

$$\bar{\sigma}_{31}^{(\alpha\beta\gamma)} = \bar{\sigma}_{31}^{(\alpha\beta\hat{\gamma})} \quad \alpha = N_\alpha/2, \beta = 1,...,N_\beta/2, \gamma = 1,...,N_\gamma/2-1$$

$$\bar{\sigma}_{12}^{(\alpha\beta\gamma)} = \bar{\sigma}_{12}^{(\hat{\alpha}\beta\gamma)} \quad \alpha = 1,...,N_\alpha/2-1, \beta = 1,...,N_\beta/2, \gamma = N_\gamma/2$$

$$\bar{\sigma}_{21}^{(\alpha\beta\gamma)} = \bar{\sigma}_{21}^{(\alpha\hat{\beta}\gamma)} \quad \alpha = N_\alpha/2-1, \beta = 1,...,N_\beta/2-1, \gamma = N_\gamma/2.$$

By rewriting the subcell stresses in terms of the subcell strains and the constitutive law, i.e.,

$$C_{11kl}^{(\alpha\beta\gamma)}\left(\bar{\varepsilon}_{kl}^{(\alpha\beta\gamma)} - \bar{\varepsilon}_{kl}^{I(\alpha\beta\gamma)} - \bar{\varepsilon}_{kl}^{T(\alpha\beta\gamma)}\right) - \quad (39)$$

$$C_{11kl}^{(\hat{\alpha}\beta\gamma)}\left(\bar{\varepsilon}_{kl}^{(\hat{\alpha}\beta\gamma)} - \bar{\varepsilon}_{kl}^{I(\hat{\alpha}\beta\gamma)} - \bar{\varepsilon}_{kl}^{T(\hat{\alpha}\beta\gamma)}\right) = 0$$

these conditions can be cast into matrix form as

$$A_M\left(\varepsilon_s - \varepsilon_s^I - \varepsilon_s^T\right) = 0. \quad (40)$$

Combining the interfacial displacement (Eq. 30) and traction (Eq. 39) conditions yields

$$\tilde{A}\varepsilon_s - \tilde{D}\left(\varepsilon_s^I + \varepsilon_s^T\right) = K\varepsilon, \quad (41)$$

where

$$\tilde{A} = \begin{bmatrix} A_M \\ A_G \end{bmatrix}, \quad \tilde{D} = \begin{bmatrix} A_M \\ 0 \end{bmatrix}, \quad K = \begin{bmatrix} 0 \\ J \end{bmatrix}. \quad (42)$$

Solving for the local subcell strains results in the final micromechanical relationship,

$$\varepsilon_s = A\bar{\varepsilon} + D\left(\varepsilon_s^I + \varepsilon_s^T\right) \quad (43)$$

where

$$A = \tilde{A}^{-1}K, \quad D = \tilde{A}^{-1}\tilde{D}. \quad (44)$$

These concentration matrices can be further decomposed into submatrices resulting in

$$A = \begin{bmatrix} A^{(111)} \\ \vdots \\ A^{(N_\alpha/2 N_\beta/2 N_\gamma/2)} \end{bmatrix}, \quad D = \begin{bmatrix} D^{(111)} \\ \vdots \\ D^{(N_\alpha/2 N_\beta/2 N_\gamma/2)} \end{bmatrix} \tag{45}$$

and leading to a relationship between the local subcell strains and globally applied strains,

$$\varepsilon^{(\alpha\beta\gamma)} = A^{(\alpha\beta\gamma)}\overline{\varepsilon} + D^{(\alpha\beta\gamma)}\left(\varepsilon_s^I + \varepsilon_s^T\right). \tag{46}$$

Lastly, the local stress in a subcell can be computed by

$$\sigma^{(\alpha\beta\gamma)} = C^{(\alpha\beta\gamma)}\left(A^{(\alpha\beta\gamma)}\overline{\varepsilon} + D^{(\alpha\beta\gamma)}\left(\varepsilon_s^I + \varepsilon_s^T\right) - \left(\varepsilon^{(\alpha\beta\gamma)I} + \varepsilon^{(\alpha\beta\gamma)T}\right)\right) \tag{47}$$

and the effective composite stress can be computed as

$$\overline{\sigma} = C^*\left(\overline{\varepsilon} - \overline{\varepsilon}^I - \overline{\varepsilon}^T\right), \tag{48}$$

where

$$C^* = \frac{8}{dhl}\sum_{\alpha=1}^{N_\alpha/2}\sum_{\beta=1}^{N_\beta/2}\sum_{\gamma=1}^{N_\gamma/2} d_\alpha h_\beta l_\gamma C^{(\alpha\beta\gamma)} A^{(\alpha\beta\gamma)} \tag{49}$$

and

$$\overline{\varepsilon}^I = -\frac{8 C^{*-1}}{dhl}\sum_{\alpha=1}^{N_\alpha/2}\sum_{\beta=1}^{N_\beta/2}\sum_{\gamma=1}^{N_\gamma/2} d_\alpha h_\beta l_\gamma C^{(\alpha\beta\gamma)}\left(D^{(\alpha\beta\gamma)}\varepsilon_s^I - \overline{\varepsilon}^{I(\alpha\beta\gamma)}\right) \tag{50}$$

and

$$\overline{\varepsilon}^T = -\frac{8 C^{*-1}}{dhl}\sum_{\alpha=1}^{N_\alpha/2}\sum_{\beta=1}^{N_\beta/2}\sum_{\gamma=1}^{N_\gamma/2} d_\alpha h_\beta l_\gamma C^{(\alpha\beta\gamma)}\left(D^{(\alpha\beta\gamma)}\varepsilon_s^T - \overline{\varepsilon}^{T(\alpha\beta\gamma)}\right). \tag{51}$$

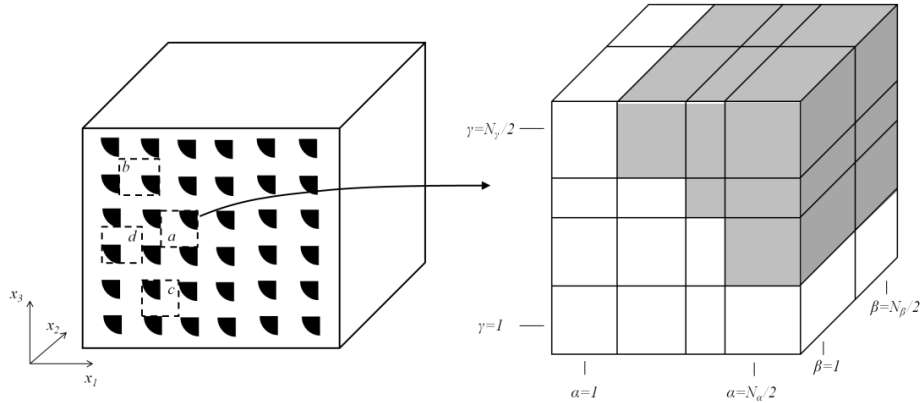

Figure 2. A repeating unit cell of a quarter circular fiber in a periodic array is shown discretized into orthogonal cuboid subcells. The dimensions of the subcells in the repeating unit cell are equivalent to that of Figure 1. Four possible, out of infinite, repeating unit cells are shown labeled *a*, *b*, *c* and *d*.

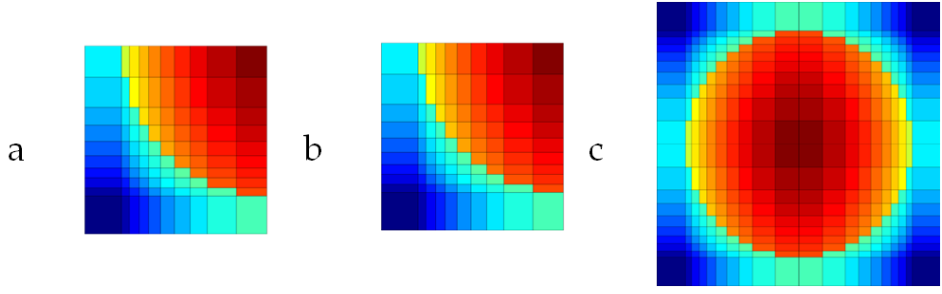

Figure 3. The ¼ symmetric theory (a), a ¼ fiber repeating unit cell (b), and a full fiber (c) all have the same local and global stress fields.

Now, let us prove that the original GMC formulation is inherently symmetric. Consider a microstructure whose periodic repeating unit cell is an eighth of the one shown in Figure 1. This is illustrated in Figure 2. This periodic unit cell is of size $d/2 \times h/2 \times l/2$ and is discretized into $N_\alpha/2 \times N_\beta/2 \times N_\gamma/2$ each of size $d_\alpha \times h_\beta \times l_\gamma$, where $d$, $h$, $l$, $N_\alpha$, $N_\beta$, and $N_\gamma$ are equivalent to that in Figure 1. The original GMC formation is summarized through Equations 41-46 and 51-59 found in Aboudi (1995). By substituting the appropriate microstructure terms into the original equations, we arrive at the following displacement continuity equations,

$$2\sum_{\alpha=1}^{N_\alpha/2} d_\alpha \bar{\varepsilon}_{11}^{(\alpha\beta\gamma)} = d\bar{\varepsilon}_{11} \qquad \beta=1,...,N_\beta/2, \gamma=1,...,N_\gamma/2 \tag{52}$$

$$2\sum_{\beta=1}^{N_\beta/2} h_\beta \bar{\varepsilon}_{22}^{(\alpha\beta\gamma)} = h\bar{\varepsilon}_{22} \qquad \alpha=1,...,N_\alpha/2, \gamma=1,...,N_\gamma/2$$

$$2\sum_{\gamma=1}^{N_\gamma/2} l_\gamma \bar{\varepsilon}_{33}^{(\alpha\beta\gamma)} = l\bar{\varepsilon}_{33} \qquad \alpha=1,...,N_\alpha/2, \beta=1,...,N_\beta/2$$

$$4\sum_{\alpha=1}^{N_\alpha/2}\sum_{\beta=1}^{N_\beta/2} d_\alpha h_\beta \bar{\varepsilon}_{12}^{(\alpha\beta\gamma)} = dh\bar{\varepsilon}_{12} \qquad \gamma=1,...,N_\gamma/2$$

$$4\sum_{\beta=1}^{N_\beta/2}\sum_{\gamma=1}^{N_\gamma/2} h_\beta l_\gamma \bar{\varepsilon}_{23}^{(\alpha\beta\gamma)} = hl\bar{\varepsilon}_{23} \qquad \alpha=1,...,N_\alpha/2$$

$$4\sum_{\alpha=1}^{N_\alpha/2}\sum_{\gamma=1}^{N_\gamma/2} d_\alpha l_\gamma \bar{\varepsilon}_{13}^{(\alpha\beta\gamma)} = dl\bar{\varepsilon}_{13} \qquad \beta=1,...,N_\beta/2.$$

and the following traction continuity conditions

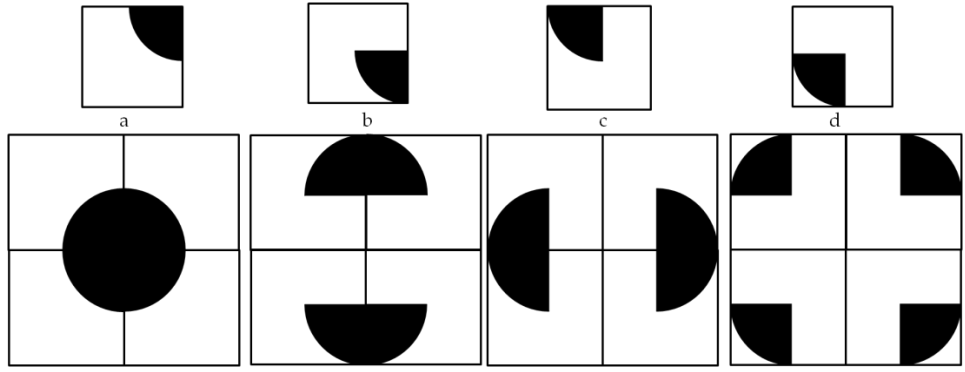

Figure 4. Due to the inherent symmetry, the four repeating unit cells of Figure 2 may actually be representative of the microstructures shown here. The original repeating unit cells produce a different microstructure than intended (quarter fiber) with the inherent symmetry conditions. The new repeating cells do not change microstructure with inherent symmetry boundary conditions.

$$\bar{\sigma}_{11}^{(\alpha\beta\gamma)} = \bar{\sigma}_{11}^{(\hat{\alpha}\beta\gamma)} \quad \alpha = 1,...,N_\alpha/2-1, \beta = 1,...,N_\beta/2, \gamma = 1,...,N_\gamma/2 \quad (53)$$

$$\bar{\sigma}_{22}^{(\alpha\beta\gamma)} = \bar{\sigma}_{22}^{(\alpha\hat{\beta}\gamma)} \quad \alpha = 1,...,N_\alpha/2, \beta = 1,...,N_\beta/2-1, \gamma = 1,...,N_\gamma/2$$

$$\bar{\sigma}_{33}^{(\alpha\beta\gamma)} = \bar{\sigma}_{33}^{(\alpha\beta\hat{\gamma})} \quad \alpha = 1,...,N_\alpha/2, \beta = 1,...,N_\beta/2, \gamma = 1,...,N_\gamma/2-1$$

$$\bar{\sigma}_{23}^{(\alpha\beta\gamma)} = \bar{\sigma}_{23}^{(\alpha\hat{\beta}\gamma)} \quad \alpha = 1,...,N_\alpha/2, \beta = 1,...,N_\beta/2-1, \gamma = 1,...,N_\gamma/2$$

$$\bar{\sigma}_{32}^{(\alpha\beta\gamma)} = \bar{\sigma}_{32}^{(\alpha\beta\hat{\gamma})} \quad \alpha = 1,...,N_\alpha/2, \beta = N_\beta/2, \gamma = 1,...,N_\gamma/2-1$$

$$\bar{\sigma}_{13}^{(\alpha\beta\gamma)} = \bar{\sigma}_{13}^{(\hat{\alpha}\beta\gamma)} \quad \alpha = 1,...,N_\alpha/2-1, \beta = 1,...,N_\beta/2, \gamma = 1,...,N_\gamma/2$$

$$\bar{\sigma}_{31}^{(\alpha\beta\gamma)} = \bar{\sigma}_{31}^{(\alpha\beta\hat{\gamma})} \quad \alpha = N_\alpha/2, \beta = 1,...,N_\beta/2, \gamma = 1,...,N_\gamma/2-1$$

$$\bar{\sigma}_{12}^{(\alpha\beta\gamma)} = \bar{\sigma}_{12}^{(\hat{\alpha}\beta\gamma)} \quad \alpha = 1,...,N_\alpha/2-1, \beta = 1,...,N_\beta/2, \gamma = N_\gamma/2$$

$$\bar{\sigma}_{21}^{(\alpha\beta\gamma)} = \bar{\sigma}_{21}^{(\alpha\hat{\beta}\gamma)} \quad \alpha = N_\alpha/2-1, \beta = 1,...,N_\beta/2-1, \gamma = N_\gamma/2.$$

By comparison of Equations 29 and 38 to Equations 51 and 52 it is apparent that the governing equations are in fact identical, thus proving that there is an inherent symmetry in GMC.

## Discussion

We numerically validate the symmetric formulation by comparing to the original formulation and then demonstrating the inherent symmetry in GMC. The numerical simulations are not intended as an all-inclusive proof, which was shown analytically, but rather as a supplemental visual verification. Consider the case of a 33% fiber volume fraction graphite epoxy composite, whose mechanical properties are listed in Table 1. The doubly periodic repeating unit cell was subjected to a transverse applied strain and Figure 3 shows the second invariant for the deviatoric stress tensor, $J_2$, for the cases of quarter symmetric GMC (Figure 3a), GMC modeling a ¼ microstructure (Figure 3b), and original GMC (Figure 3c). It is qualitatively clear that the local fields are equivalent for all three cases. The global fields (stress, strain, and stiffness) are also equivalent. This implies that existing GMC theories and their derivatives need only represent a fourth (doubly periodic) or eighth (triply periodic) of the RUC, this also implies that there is no benefit for modeling the entire RUC.

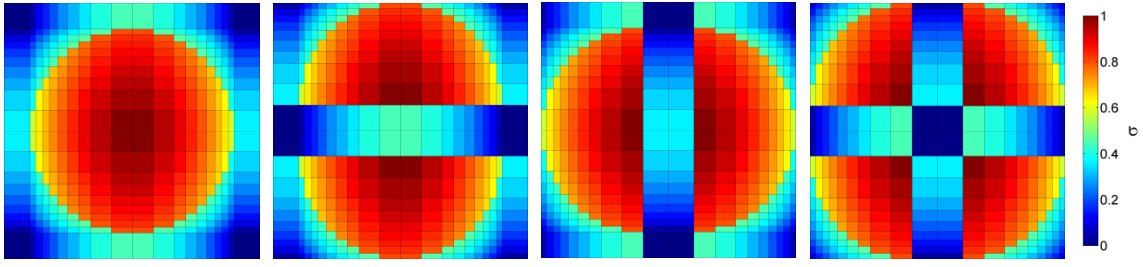

Figure 5. The effective stress fields (normalized to the maximum) are equivalent for all microstructures shown in Figure 4. This means that each subcell will have the same stress levels to some extent regardless of ordering. This implies that GMC can only simulate continuous and ordered microstructures due to the lack of shear coupling. The global elastic constants are shown in Table 2.

While the number of unknowns are generally reduced by ¼ or 1/8 for the doubly or triply periodic case, respectively, as with the case of the method of cells there is a critical size before the increased efficiency become realizable. For modeling circular fibers, the most efficient representation is an even number of subcells in the beta and gamma directions, fiber positioned in a corner and with its edges touching the RUC boundaries. Say the fiber is discretized into an odd number of subcells, $n$, across its width then the total number of subcell is $(n+1)^2$. A quarter symmetric RUC requires $((n+1)/2+1)^2$ subcells. Only for the case of a 2x2 repeating unit cell (one subcell for the fiber), are the numbers of unknowns equivalent. As the fiber becomes more refined, the unknowns decrease with $1/(n+1)^2$. For a triply periodic RUC, it is $1/(n+1)^3$.

Next, we explore the implications of the inherent symmetry. The classical method of cells (Aboudi, 1989) repeating a unit cell can be thought of as a quarter symmetric representation. The classical square fiber RUC and a quarter square fiber RUC are identical in relative dimensions. Eq. 29 implies that the results are independent of absolute dimensions ($d$, $h$, and $l$) and are only dependent on relative dimensions. Therefore, the classic method of cell RUC, the square fiber in a square array, is actually a quarter square fiber in square array due to the inherent symmetry. This is interesting because the microstructures for the entire RUC and quarter symmetric RUC are indistinguishable, both requiring $N_b$, $N_g=2$ of the same relative size. In fact, scaling the dimensions of the RUC may improve the condition of the concentration matrix in Eq. 41. The same conclusion holds to a cubic inclusion in an ordered array, whereas an eighth and full representation both require $N_a$, $N_b$, $N_g=2$.

Table 1. Constituent Elastic Properties of Graphite/Epoxy Material System

|  | Axial Modulus (GPa) | Transverse Modulus (GPa) | Axial Poisson's Ratio | Transverse Poisson's Ratio | Shear Modulus (GPa) |
| --- | --- | --- | --- | --- | --- |
| Graphite | 230.0 | 15.0 | 0.2 | 0.2 | 15.0 |
| Epoxy | 3.1 | 3.1 | 0.38 | 0.38 | 1.12 |

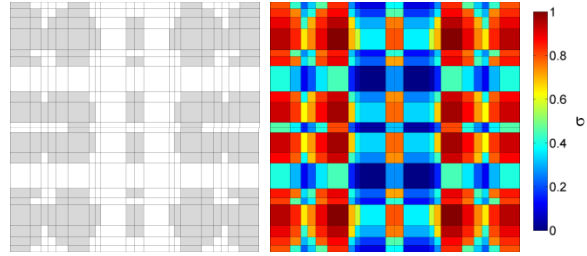

Figure 6. Random permutations (a) of the full circular fiber repeating unit cell in Figure 3a show that the effective stress fields (b) do not change. This implies that as long as each row and column in a repeating unit cell (also in triply periodic) has the same relative volume fractions as another microstructure, regardless or ordering, the local and global fields will be equivalent.

Due to the inherent symmetry, there may be periodic unit cells which GMC cannot accurately model. Take the case of a quarter fiber (Figure 2). There are an infinite number of doubly periodic repeating unit cell configurations possible, but we focus on the four shown in Figure 2a. All four RUCs provide the same response using GMC. The first ambiguity arises when consider Figure 4a in the quarter symmetry context, is actually representative of a circular fiber. This leads to the conclusion that GMC cannot distinguish between a ¼ fiber and whole fiber. By studying the remaining RUCs, we can understand the other ambiguities that arise. Figures 4b-d show the full representation of the inherent ¼ symmetry, all of which represent drastically different microstructures. To numerically validate this, all repeating unit cells were run using the original GMC theory. The stress field is shown in Figure 4 and the moduli are reported in Table 2. Both local and global fields are identical is due to the lack of shear coupling in GMC, a facet the High Fidelity Generalized Method of Cells (HFGMC) (Aboudi, Arnold, and Bednarcyk, 2012) overcomes. We also present a simple method to reason this: consider the "square" fiber RUC (one subcell for the fiber). Due to the average enforcement of boundary conditions and lack of shear coupling, a "square" fiber RUC is representative of a circular fiber. With this mindset, a quarter of a square, a smaller square, in actuality is a quarter circle. If the quarter circle in Figures 5b-5d is substituted with a square fiber, then it is acceptable to reason that all five RUCs produce the same periodic unit cell. The fundamental observation is GMC is insensitive to the positioning and orientation of inclusions, as long as the total array exhibits order. For example, we can randomly swap and rows or columns and receive the same global and local response, which is shown in Figure 6. This leads to the conclusions that regardless of order, as long as each row and column has the same relative volume fractions as another microstructure, the global and local fields will be identical. This is another explanation supporting the inherent symmetry. This occurs due to the isostress and isostrain conditions that arise from the first-order theory. This also has an interesting implication: consider a high resolution square fiber. Rearranging the rows and columns can produce a smaller RUC of multiple square fibers (Figure 7), both of which are identical in global and local response. With these realizations, when a complex microstructure is simulated one must question which other microstructures may also being simulated.

Table 2. Effective Elastic Constants of Various Microstructures

|  | Axial Modulus (GPa) | Transverse Modulus (GPa) | Axial Poisson's Ratio | Transverse Poisson's Ratio | Shear Modulus (GPa) |
|---|---|---|---|---|---|
| ¼ Fiber | 116.7 | 7.67 | 0.293 | 0.41 | 2.06 |
| Config A | 116.7 | 7.67 | 0.293 | 0.41 | 2.06 |
| Config B | 116.7 | 7.67 | 0.293 | 0.41 | 2.06 |
| Config C | 116.7 | 7.67 | 0.293 | 0.41 | 2.06 |
| Config D | 116.7 | 7.67 | 0.293 | 0.41 | 2.06 |
| Random Permutation | 116.7 | 7.67 | 0.293 | 0.41 | 2.06 |

The intent of this paper is not to cast doubt on the accuracy or reliability of GMC, but to illustrate a facet of all first-order micromechanical theories. More specifically, this paper is intended to raise awareness that global solutions are not unique to repeating unit cells and repeating unit cells may not be representative of the microstructure. Lastly, we recommend the following considerations when selecting a micromechanical model. Mean field theories and zeroth-order micromechanical methods cannot accommodate architecture. First-order theories have ambiguity in their architecture. Therefore, for complex architectures, higher-order theories such as HFGMC, are necessary to capture these effects.

## Conclusion

The original intent of this paper was to explore symmetric boundary conditions and increase the efficiency of the GMC. Symmetric boundary conditions were applied in the continuity of traction and continuity of displacement governing equations. In the process, we discovered an inherent symmetry in GMC, negating the need for a symmetric formulation. While exploring this inherent symmetry, we demonstrated that first-order theories are insensitive to the position of constituent subcells. We showed that randomly swapping rows and columns in a periodic fiber repeating unit cell produces the same global and local fields. This implies that regardless of ordering, if two microstructures have rows and columns with identical relative volume fractions, there global and local responses will be the same. The ambiguity and insensitivity arises from the lack of shear coupling and the isostrain and isostress conditions that result from first-order displacement field assumptions. We also concluded that first-order theories are best suited for regular or ordered microstructures, and to capture complex microstructures, higher-order theories are necessary.

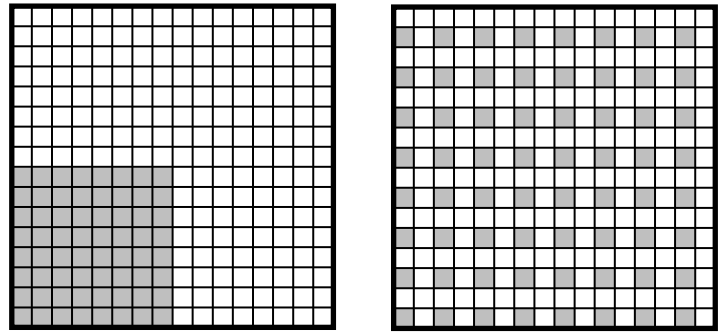

Figure 7. There is no mesh dependency in the GMC; therefore, the response of the left microstructure will not change with refinement. Interestingly, the microstructure on the right can be created from swapping rows and columns of the left microstructure and the result is the same repeating unit cell. This microstructure maintains the same relative volume fractions in each row and column (ordering is not important).

## Acknowledgements

This research was supported in part by an appointment to the Postgraduate Research Participation Program at the U.S. Army Research Laboratory administered by the Oak Ridge Institute for Science and Education through an interagency agreement between the U.S. Department of Energy and USARL.